\documentstyle[aps]{revtex}
\begin{document}
 \draft
\def\be{\begin{equation}}
\def\ee{\end{equation}}
\def\ba{\begin{eqnarray}}
\def\ea{\end{eqnarray}}
\def\bq{\begin{quote}}
\def\eq{\end{quote}}
\def\PL{{ \it Phys. Lett.} }
\def\PRL{{\it Phys. Rev. Lett.} }
\def\NP{{\it Nucl. Phys.} }
\def\PR{{\it Phys. Rev.} }
\def\MPL{{\it Mod. Phys. Lett.} }
\def\IJMP{{\it Int. J. Mod .Phys.} }
\newcommand{\labell}[1]{\label{#1}\qquad_{#1}} 
\newcommand{\labels}[1]{\vskip-2ex$_{#1}$\label{#1}} 

\twocolumn[\hsize\textwidth\columnwidth\hsize\csname
@twocolumnfalse\endcsname
\preprint{LBNL-44030\\NYU-TH-/99/07/02\\SU-ITP-99/36\\
hep-th/9907209 \\ July 1999}
\date{\today}
\title{Infinitely Large New Dimensions}
\author{Nima Arkani-Hamed$^{1,2}$, Savas Dimopoulos$^{3}$, Gia
Dvali$^{4}$ and Nemanja Kaloper$^{3}$}
\vskip.5cm
\address{{\em $^1$
Department of Physics, University of California, Berkeley, CA 94530, USA}\\
{\em $^2$Theory Group, Lawrence Berkeley National Laboratory,
Berkeley, CA 94530, USA}\\
{\em $^3$
Department of Physics, Stanford University, Stanford, CA 94305, USA}\\
{\em $^4$ Department of Physics, New York
University, New York, NY 10003, USA and ICTP, Trieste, Italy}\\}
\maketitle
\begin{abstract}
We construct intersecting brane configurations in Anti-de-Sitter space
localizing gravity to the intersection region, with any number $n$ of
extra dimensions. This allows us to construct two kinds of theories
with infinitely large new dimensions, TeV scale quantum gravity and
sub-millimeter deviations from Newton's Law. The effective 4D Planck scale
$M_{Pl}$ is determined in terms of the fundamental Planck scale $M_*$
and the $AdS$ radius of curvature $L$ via the familiar relation
$M_{Pl}^2 \sim M_{*}^{2+n} L^n$; $L$ acts as an effective radius of
compactification for gravity on the intersection. Taking $M_* \sim $ TeV
and $L \sim $ sub-mm reproduces the phenomenology of theories with large
extra dimensions. Alternately, taking $M_* \sim L^{-1} \sim M_{Pl}$, and
placing our 3-brane a distance $\sim 100 M_{Pl}^{-1}$ away from the
intersection gives us a theory with an exponential determination of the
Weak/Planck hierarchy.
\end{abstract}
\pacs{PACS:12.10.-g,11.10.Kk, \hfill  hep-th/9907209, LBNL-44030,\\
11.25.M,04.50.+h \hfill  NYU-TH/99/07/02, SU-ITP-99/36}
\vskip2pc]

Unification of gravity with other forces of nature
suggests that the world has more than three spatial
dimensions. Since only three of these
are presently observable,
one has to explain why the additional ones have
eluded detection. The conventional explanation is that
the dimensions are compactified with tiny radii of order the
Planck length $\sim 10^{-33}$ cm, which makes them impossible to
probe with currently available energies.

It has recently been pointed out that new dimensions may have
a size $R$  much larger
than the fundamental Planck length of the theory, perhaps as
large as a millimeter \cite{savas,ahdmr}.
This has the effect of diluting the strength of the $4D$ gravity observed
at distances much larger than $R$. The $4D$ Planck scale $M_{Pl}$
is determined by the fundamental Planck scale $M_*$ via Gauss's law
$M_{Pl}^2 \sim M_{*}^{2+n} R^n$ where $n$ is the number of new dimensions.
The original motivation was to bring
the fundamental gravitational scale close to the weak scale in order to
solve the hierarchy problem. These large dimensions are not in conflict with
experiment if the Standard Model fields are confined to a
3-brane in the extra
dimensions.

In this scenario, the only reason to compactify the extra
dimensions at all is to reproduce $4D$ Newtonian gravity at long distances.
One can wonder if even this is necessary: if gravity itself is somehow
``trapped'' to our 3-brane, then $4D$ gravity can be reproduced even if
the extra dimensions are infinitely large. A very interesting
recent construction by
Randall and Sundrum \cite{rs} and by Gogberashvili \cite{gog} provides an
explicit realization of this idea for the case of one extra dimension.
Solving Einstein's equations with a $3$-brane in $(4+1)$ dimensions,
together with a bulk
cosmological constant, they find a massless $4D$ graviton localized
to the 3-brane.
The resulting gravitational potential between
any two objects on the brane is inversely
proportional to the distance between the objects, and
not its square, despite the presence of the
infinite fifth dimension.

It is clearly desirable to extend this idea to any number of new dimensions.
At first sight, however, the mechanism of \cite{rs,gog} seems to rely
on the peculiar properties of co-dimension one objects in gravity and
seems hard to extend to the case of more dimensions.
However, all that
seems to be required
is the presence of {\it some}
co-dimension one branes in the system, while our
$3$-brane
can have larger co-dimension. We are
led to consider a system of $n$ mutually intersecting $(2+n)$ branes
in $(3+n) + 1$ dimensions with a bulk cosmological constant.
The branes intersect on 3 spatial dimensions, where the
Standard Model fields
reside. Intuitively, each of the
$(2+n)$ branes has co-dimension one and tries to localize gravity to
itself. Therefore gravity will be localized to the intersection of all
the branes! We will now confirm this intuition by explicit calculations.

We begin by deriving the solution describing the intersection
of branes. Consider an array of $n$ orthogonal
$n+2$-spatial dimensional branes in $(3+n) + 1$ dimensions,
with a bulk cosmological constant $\Lambda$.
For simplicity we take the branes to have identical tension $\sigma$.
The field equations can be
derived from the action
\ba
S &=& \int_M d^{4+n} x \sqrt{g_{4+n}}
\Bigl(\frac{1}{2\kappa^2_{4+n}} R + \Lambda\Bigr) \nonumber \\
&&~~~~~~~~~~~
 - \sum_{k=1}^n \int_{\mbox{k'th brane}} d^{3+n} x \sqrt{g_{3+n}} \sigma .
\label{action}
\ea
Here $\kappa^2_{4+n} = 8\pi/M^{n+2}_*$, where $M_*$ is the
fundamental scale of the theory.
Note that the measure of integration differs between each brane,
and between the branes and the bulk. This will be reflected
in the field equations, where ratios
$\frac{\sqrt{g_{3+n}}}{\sqrt{g_{4+n}}}$
weigh the $\delta$-function sources. After the standard Euler-Lagrange
variational procedure, the field equations are
\ba
&&R^a{}_b - \frac12 \delta^a{}_b R =
\kappa^2_{4+n} \Lambda \delta^a{}_b
\nonumber \\
&&~~~ - \frac{\sqrt{g_{3+n}(1)}}{\sqrt{g_{4+n}}}
\kappa^2_{4+n} \sigma \delta({\bar z}^1)
{\rm diag} (1,1,1,1,0,...,1,1) \nonumber \\
&&~~~ - ...  \nonumber \\
&&~~~ - \frac{\sqrt{g_{3+n}(n)}}{\sqrt{g_{4+n}}}
\kappa^2_{4+n} \sigma \delta({\bar z}^n)
{\rm diag}(1,1,1,1,1,1,...,0),
\label{eoms}
\ea
where the coordinates ${\bar z}^k$ parameterize the extra dimensions.
We note that the ratios $\frac{\sqrt{g_{3+n}(k)}}{\sqrt{g_{4+n}}}$
reduce to $\sqrt{g^{kk}}$ for diagonal metrics. In general, they
cannot be gauged away.

It is now straightforward to
write down the solutions. Away from the branes,
the solution in the bulk comprises of patches of
the $4+n$-dimensional Anti-de-Sitter space.
Hence if the branes are mutually orthogonal,
by symmetry the full solution simply consists of $2^{n}$ identical
patches of the $AdS_{n+4}$ which fill up the
higher-dimensional quadrants between the branes, and are glued together
along the branes.
To construct it, we
start with the Poincare half-plane parametrization of $AdS_{n+4}$,
given by
\be
ds^2_{n+4} = \frac{L^2}{z^2}
\Bigl(\eta_{\mu\nu} dx^\mu dx^\nu + d\vec w_{n-1}^2 + dz^2\Bigr).
\label{ads}
\ee
The length scale $L$ is determined by the bulk cosmological constant as
\be
L^2 = \frac{(n+3)(n+2)}{2\kappa^2_{4+n} \Lambda}.
\label{bulklength}
\ee
To make use of the symmetry, it is convenient
to find the patch of (\ref{ads}) where the metric is
manifestly symmetric under permutations of all extra
dimensions. This is most easily accomplished
by an $O(n)$
rotation of $\vec w_{n-1}, z$. We transform to new coordinates
${\bar z}^k$, $k \in \{1, ...,n\}$ by a rigid rotation
chosen such that $z = \sum^n_{j=1} {\bar z}^j/\sqrt{n}$. In terms of these
coordinates, the metric is
\be
ds^2_{n+4} = \frac{n L^2}{(\sum^n_{j=1} {\bar z}^j)^2}
\Bigl(\eta_{\mu\nu} dx^\mu dx^\nu + \sum^n_{k=1} (d{\bar z}^k)^2 \Bigr).
\label{adsnew}
\ee
The metric (\ref{adsnew}) covers a segment of the extra dimensions
bounded by the branes.
Led by our discussion above, we will
take such a cell of $AdS_{n+4}$ with e.g. all the $\bar{z}^j > l$
for some $l$. We then fill
out the rest of the space by reflecting this cell in all $2^n$ distinct ways
about its boundaries. The resulting metric is given by replacing
$\sum_j \bar{z}^j$ by $\sum_j |\bar{z}^j| + l$ in eqn.(\ref{adsnew}).
By rescaling $x,\bar{z}$, we can set $l$ to any value we wish, and we
use this freedom to put the metric in the final form
\footnote{
For $n=1$, the solution coincides with that given by
\cite{rs,gog} if we make the coordinate
transformation $L \exp(|y|/L)=|z|+ L $, and $k = 1/L$.}
\be
ds^2_{n+4} = \frac{1}{(k\sum^n_{j=1} |{\bar z}^j| +  1)^2}
\Bigl(\eta_{\mu\nu} dx^\mu dx^\nu + \sum^n_{k=1} (d{\bar z}^k)^2 \Bigr),
\label{adsfinal}
\ee
where $k \equiv (\sqrt{n} L)^{-1}$.
This choice corresponds to setting the conformal factor in
(\ref{adsfinal}) to unity at the intersection ${\bar z}^k = 0$.
Physically this means that the unit of length on the
intersection is set by $M^{-1}_*$.
In this metric,
each ${\bar z}^j$ is allowed to vary on the whole real line.
The curvature of the space will now have singularities at the seams
where we have pasted together
the elementary cells, but these will be precisely those
dictated by the presence of the branes.

It is straightforward to see this explicitly. The metric is a conformal
transformation of flat space $g_{ab} = \Omega^2 \eta_{ab}$ where
\be
\Omega = \frac{1}{k\sum_j |\bar{z}^j| + 1}
\ee
is the ``warp factor''.
We can trivially compute the Einstein tensor $G_{ab} = R_{ab} - 1/2 g_{ab} R$
using the standard relation (in general for $g_{ab} = \Omega^2
\tilde{g}_{ab}$ for $D$ spacetime dimensions)
\ba
G_{ab} &=& \tilde{G}_{ab} + (D-2) \left( \tilde{\nabla}_a \mbox{log} \Omega
\tilde{\nabla}_b \mbox{log} \Omega - \tilde{\nabla}_a \tilde{\nabla}_b
\mbox{log} \Omega\right) \nonumber  \\
&+& (D-2)\tilde{g}_{ab} \left(\tilde{\nabla}^2 \mbox{log} \Omega
+ \frac{D-3}{2} (\tilde{\nabla} \mbox{log} \Omega)^2 \right).
\label{conformal}
\ea
Using this it is easy to compute the Einstein tensor for our metric, and we
find
\ba
&& G^{a}{}_b = \kappa^2_{4+n} \frac{n(n+2)(n+3) k^2}{2}
\delta^a{}_b \nonumber \\
&& ~~~~ -\kappa^2_{4+n}
\frac{2(n+2) k}{\Omega} \delta(\bar{z}^1) (1,1,1,1,0,1,\cdots,1)
\nonumber\\ && ~~~~ - \cdots \nonumber \\
&& ~~~~ -\kappa^2_{4+n}
\frac{2(n+2) k}{\Omega} \delta(\bar{z}^n) (1,1,1,1,1,1,\cdots,0),
\ea
which reproduces eqn.(2) if the brane tension $\sigma$ is chosen to be
$\sigma = 2(n+2) k$.

In order to demonstrate that we have
indeed localized gravity to the intersection,
we must look at the linear perturbations
about this solution. It is convenient
to parameterize the perturbations by
replacing $\eta_{\mu \nu}$ with $\eta_{\mu \nu}
+ h_{\mu \nu}(x,\bar{z})$ in eqn.(\ref{adsfinal}).
Again using the conformal
transformation eqn.(\ref{conformal}) we easily find the linearized field
equations for $h_{\mu \nu}$, which are in the gauge $h^{\mu}{}_{\mu} =0,
\partial_{\alpha} h^{\alpha\mu} = 0$
\be
\left[\Box_4 - \nabla^2 _{\bar{z}} + (n+2) \Omega \sum_j
\mbox{sgn}(\bar{z}^j) \partial_{j} \right] h(x,\bar{z}) = 0,
\ee
where we have dropped the $\mu \nu$ index on $h$.
The transverse-traceless gauge is invariant under
conformal transformations on the background (\ref{adsfinal}),
and hence our calculation in the conformal frame
exactly reproduces the results in the original frame (\ref{adsfinal}).
This immediately shows that
the ordinary four-dimensional graviton is present as a massless mode in the
theory, corresponding to a $\bar{z}$
independent solution $h(x,\bar{z}) = h(x)$.
Indeed, replacing $\eta_{\mu \nu}$ by
$g^{(4)}_{\mu \nu}(x)$ in eqn.(\ref{adsfinal})
and inserting into the action, we find
\be
S = \int d^n \bar{z} \Omega^{2 + n} \times \int d^4 x \sqrt{g^{(4)}} R^{(4)},
\ee
which shows that the four-dimensional graviton couples with strength
\be
M_{Pl} ^2 = M_*^{2 + n} \int d^n \bar{z} \Omega^{2 + n} \sim M_*^{2+n} L^n.
\ee
The exact calculation gives $M^2_{Pl} = \frac{2^n n^{n/2}}{(n+1)!}
M_*^{2+n} L^n$. This relation suggests that $L$ can be interpreted as the
effective size of $n$ compact dimensions, even though the extra
dimensions are infinitely large.
This is indeed a correct interpretation as we will see shortly. For
a complete analysis of the effective $(3+1)D$ spectrum, it is convenient
to make a change of variables $h = \Omega^{-(n+2)/2} \hat{h}$, in terms
of which the linearized equations are
\be
\left[\frac{1}{2} \Box_4 + \left(-\frac{1}{2}
\nabla^2_{\bar{z}} + V(\bar{z}) \right) \right] \hat{h} = 0,
\label{poteq}
\ee
where
\be
V(\bar{z}) = \frac{n(n+2)(n+4)k^2}{8} \Omega^2
- \frac{(n+2)k}{2} \Omega \sum_j \delta(\bar{z}^j).
\label{pot}
\ee
In order to determine the spectrum of $4D$ masses,
we set $\hat{h} = e^{i p x}
\hat{\psi}(\bar{z})$; the $4D$ masses are
then determined by the eigenvalues
of an effective $n$ dimensional Schr\" odinger equation
\be
\left(-\frac{1}{2} \nabla^2_{\bar{z}} + V(\bar{z})
\right) \hat{\psi}_\lambda = \frac{1}{2} m^2_\lambda \hat{\psi}_\lambda,
\ee
where $\lambda$ labels the eigenfunctions.
All of the important physics follows
from a qualitative analysis of this potential and parallels the
story with one extra dimension. The potential has a repulsive piece which
goes to zero for $\sum_{j} |\bar{z}^j| \gg L$, and a sum of
attractive $\delta$ functions.
We already know that the $4D$ massless graviton
corresponds to a bound state with the
wavefunction (numerical factors will be
omitted in all that follows)
\be
\hat{\psi}_{\mbox{bound}} \sim \Omega^{(n+2)/2}.
\ee
Since the potential falls off to
zero at infinity, we will also have continuum
modes. Since the height of the potential near the origin is $\sim k^2$,
the modes with $m^2 < k^2$ will have suppressed
wavefunctions, while those with
$m^2 > k^2$ will sail over the potential and
will be unsuppressed at the origin.
In order to see the physics more explicitly, suppose we place a test
mass $M$ on the intersection at $(x=0,\bar{z}=0)$, and ask for the
gravitational potential $U(r)$
at a distant point on the intersection $(|x|=r,\bar{z}=0)$.  To do this,
we simply insert a source $G_{N(4+n)}
M \delta^3(x) \delta^n(\bar{z})$ on the RHS of
eqn.(\ref{poteq}), and
straightforwardly solve the equation to find
\ba
\frac{U(r)}{M} &=& \sum_\lambda G_{N(4+n)} |\hat{\psi}_\lambda(0)|^2
\frac{e^{-m_\lambda r}}{r} \nonumber \\
\sim  \frac{G_{N(4+n)}}{L^n} \frac{1}{r} &+& \sum_{\mbox{continuum}}
G_{N(4+n)} |\hat{\psi}_\lambda(0)|^2
\frac{e^{-m_\lambda r}}{r}.
\ea
In the second line we have separated the bound-state
from the continuum
contributions. It is straightforward to evaluate the suppression of
$|\psi_\lambda(0)|$ for modes with $mL <1$,
but this is not needed for the discussion of
the limiting behaviour of $U(r)$. Consider first large distances $r \gg L$.
Even with {\it no} suppression of the continuum modes for $mL < 1$, the
continuum sum would yield the $(4+n)D$ potential $G_{N(4+n)}/r^{n+1}$, which
is sub-dominant to the term generated by the $4D$ graviton bound state for
$r \gg L$. Therefore, for $r \gg L$,
\be
U(r) \sim \frac{G_{N(4)} M}{r},
~~~~~~~~ G_{N(4)} \sim \frac{G_{N(4+n)}}{L^n}.
\ee
On the other hand, for distances $r \ll L$, it is the continuum modes with
$m L \gg 1$ which dominate, and these have unsuppressed wavefunctions at
the origin. Therefore, for $r \ll L$, we just get the $(4+n)$D potential
\be
U(r) \sim \frac{G_{N(4+n)}M}{r^{n+1}}.
\ee
This point was not discussed in \cite{rs}, as they were only interested in
checking the $r \gg L$ behavior. Of course, a precise treatment is needed
to understand the details of the cross-over between these limits. However,
the qualitative behavior is exactly what we would expect by
interpreting $L$ as a ``compactification radius''. This is in accordance with
the intuition that, while the extra
dimensions are infinitely large, gravity is
localized to a region of size $L$ around the intersection of the branes.
The mechanism of localization is realized by the branes
repelling all graviton modes with bulk momentum smaller than
$1/L$ but greater than zero away from the intersection,
to distances of size $L$.
As a result, inside of this region gravity becomes weak
and the resulting Planck scale can be many orders of magnitude
larger than the fundamental scale.
The length scale $L$ is determined by the
bulk cosmological constant $\Lambda$,
and given our ignorance regarding the
cosmological constant problem, we do not
feel any strong prejudice forcing $L$ to be of the order of the fundamental
scale $M_*^{-1}$. We will simply treat $L$ as a parameter; we know only that
$L$ must be smaller than $\sim $ 1 mm from the present-day gravity
measurements.

For $L \gg M_*^{-1}$, the phenomenology seems to be
very similar to conventional large extra dimensions of size $L$, but there
is a different theoretical perspective. In particular, the problem of
stabilizing the radius at large values is replaced here with explaining
the tiny bulk cosmological constant and brane tensions. Furthermore, there
do not seem to be any very light
moduli fields associated with the large
``radius'' $L$.

Another possibility motivated by the first paper in \cite{rs} is to
stay with $L \sim M_*^{-1}$, and use the
factor $\Omega$ to exponentially generate the weak scale on our 3-brane,
which is placed a distance $O(100) M_*^{-1}$ away from the
the intersection. Unlike the proposal in the first paper of \cite{rs},
this can be done with infinitely large extra dimensions. This is because
in our case, 3-branes are tiny compared to the $(2+n)$ branes setting up the
gravitational background, so they are just like test particles probing the
background geometry. Such a ``hybrid'' model has interesting
phenomenology. Since the bulk is infinitely large, in contrast to the
first paper in \cite{rs} there is a continuum of graviton modes, and they
lead to a correction to the Newtonian potential on our
brane $\sim ($TeV$)^{-(n+2)} r^{-(n+1)}$ at
all distances. This correction is irrelevant at large distances, but
dominates the Newtonian potential at distances smaller than $R$ where
$M_{Pl}^2 \sim ($TeV$)^{n+2} R^n$. This framework combines the interesting
features of having infinitely large new dimensions, exponential
determination of the Weak/Planck hierarchy, strong gravity at the TeV
scale and possible sub-millimeter deviations from Newtonian gravity.

In sum, we have shown that gravity can be
localized to the intersection of orthogonal $(2+n)$ branes lying in
infinite $AdS_{n+4}$ space.
Our solution naturally generalizes the example with one extra dimension of
\cite{rs,gog}.
It is therefore possible to mask any number of infinitely large
extra dimensions.
Furthermore, we pointed out that the curvature
$L$ of the bulk $AdS$ space acts as an effective ``compactification'' scale;
the Newtonian potential on the intersection behaves as $1/r$ for $r \gg L$
and $1/r^{n+1}$ for $r \ll L$. Among other things, this could offer new
possibilities for constructing theories with sub-millimeter extra dimensions.

A number of aspects of our set-up need to be further elaborated.
For instance, in the construction we have assumed that there is no extra
tension localized at the intersection
of the branes, e.g. we have not included
possible 3-brane sources of the form $\delta^n(\bar{z})$ in Einstein's
equations. It is important to determine how our solutions are modified
in the presence of such sources. There are obvious generalizations
to intersecting branes with different tensions or at general angles to each
other, and we expect that these will mimic the
physics of anisotropic compactification.
There are many fascinating questions
left to ask, particularly in cosmology
where interesting new issues and possibilities have already emerged
in the context of sub-millimeter dimensions\cite{kl,dt,ahdkmr}.
It is also interesting to
explore the extent to which the
(now infinitely large) bulk can be used to
address other mysteries of the Standard Model,
perhaps along the lines of \cite{more}.
We intend to pursue these issues in future investigations.

\vspace{.5cm}
{\bf Acknowledgements}

We thank John March-Russell for discussions, and for informing us that he
has found different solutions with more than one extra dimension.
The work of N.A-H. has been supported in part by the DOE under Contract
DE-AC03-76SF00098, and in part by NSF grant PHY-95-14797.
The work of S.D. and N.K has
been supported in part by NSF Grant PHY-9870115.

\end{document}